\documentclass[a4paper]{article}
\usepackage{amsmath}
\usepackage{amsfonts}
\usepackage{amssymb}
\usepackage{amsthm}
\usepackage[dvips]{graphicx}
\usepackage{authblk}
\usepackage{cite}

\newtheorem{proposition}{Proposition}

\renewcommand\qedsymbol{$\blacksquare$}
\newcommand*{\Scale}[2][4]{\scalebox{#1}{$#2$}}

\providecommand{\keywords}[1]
{
  \small	
  \textbf{Keywords:} #1
}

\newcommand{\be}{\begin{equation}}
\newcommand{\ee}{\end{equation}}

\numberwithin{equation}{section}

\begin{document}

\title{Deformation of the Heisenberg-Weyl algebra and the Lie superalgebra $\mathfrak{osp}\left( {1|2} \right)$: exact solution for the quantum harmonic oscillator with a position-dependent mass}

\author[1]{E.I.~Jafarov\footnote{Corresponding author: ejafarov@physics.science.az}}
\author[1]{S.M.~Nagiyev\footnote{sh.nagiyev@physics.science.az}}
\author[2]{J.~Van~der~Jeugt\footnote{Joris.VanderJeugt@UGent.be}}

\affil[1]{Institute of Physics, Ministry of Science and Education \\Javid av. 131, AZ1143, Baku, Azerbaijan}

\affil[2]{Department of Applied Mathematics, Computer Science and Statistics, Ghent University \\ Krijgslaan 281-S9, 9000 Gent, Belgium}

\date{} 

\maketitle

\begin{abstract}

We propose a new deformation of the quantum harmonic oscillator Heisenberg-Weyl algebra with a parameter $a>-1$. 
This parameter is introduced through the replacement of the homogeneous mass $m_0$ in the definition of the momentum operator $\hat p_x$ as well as in the creation-annihilation operators $\hat a^\pm$ with a mass varying with position $x$. 
The realization of such a deformation is shown through the exact solution of the corresponding Schr\"odinger equation for the non-relativistic quantum harmonic oscillator within the canonical approach. 
The obtained analytical expression of the energy spectrum consists of an infinite number of equidistant levels, whereas the wavefunctions of the stationary states of the problem under construction are expressed through the Hermite polynomials. 
Then, the Heisenberg-Weyl algebra deformation is generalized to the case of the Lie superalgebra $\mathfrak{osp}\left( {1|2} \right)$. 
It is shown that the realization of such a generalized superalgebra can be performed for the parabose quantum harmonic oscillator problem, the mass of which possesses a behavior completely overlapping with the position-dependent mass of the canonically deformed harmonic oscillator problem. 
This problem is solved exactly for both even and odd stationary states. 
It is shown that the energy spectrum of the deformed parabose oscillator is still equidistant, however, both even and odd state wavefunctions are now expressed through the Laguerre polynomials. 
Some basic limit relations recovering the canonical harmonic oscillator with constant mass are also discussed briefly.

\end{abstract}


\section{Introduction}

The uncertainty principle of quantum mechanics is one of the most important foundation elements of its definition. 
The limit for simultaneous measurements of position and momentum of certain quantum systems imposed by this principle is well known. 
A simpler mathematical formulation of this principle is based on the canonical commutation relation between the position and momentum operators, which states that their commutator should result in $i \hbar$, i.e. a $c$-number~\cite{born1925,heisenberg1927,kennard1927,weyl1927}. Such a commutator is valid only if the momentum operator in the position representation is simply defined as a first-order ordinary derivative with respect to position. 
This definition of the position operator further leads to the elegant and simple analytical solutions for the wavefunctions of the stationary states and equidistant energy spectrum in the quantum-mechanical treatment of the harmonic oscillator~\cite{landau1991}. 
For example, the equidistant energy spectrum within the canonical approach is well-known:

\be
\label{e-c}
E_n=\hbar \omega \left( n + \frac 12 \right), \qquad n=0,1,2,\ldots.
\ee
The dynamical symmetry algebra of this non-relativistic oscillator problem is also well known. 
It is the Heisenberg-Weyl algebra with three generators -- the creation and annihilation operator, and the Hamiltonian itself~\cite{dirac1927,infeld1951}.

In his seminal paper~\cite{wigner1950}, Wigner opened the discussion of the unique determination of the commutation relation between the position and momentum operator. 
He proved for the problem of the quantum harmonic oscillator that the supremacy of the Heisenberg-Lie equations on the non-commuting nature of the position and momentum operator leads to a picture in which their commutator is no longer a $c$-number. 
His computations led to a non-canonical equidistant energy spectrum that generalizes \eqref{e-c} through a positive parameter $\gamma$:

\be
\label{e-nc}
E_n=\hbar \omega \left( n + \gamma \right), \qquad \gamma \geq 1/2.
\ee
This discussion of the uniqueness of the canonical commutation relations further led to the development of the second quantization concept for statistics that differs from both Bose-Einstein and Fermi-Dirac statistics~\cite{yang1951,green1953}. 
For the harmonic oscillator, it allowed more general analytical solutions governed by a dynamical symmetry Lie superalgebra $\mathfrak{osp}\left( {1|2} \right)$ in both configuration~\cite{mukunda1980,ohnuki1982} and phase space~\cite{jafarov2008}, and the system is referred to as the parabose quantum harmonic oscillator.

However, the problem highlighted above is true if we deal with a quantum oscillator system with a homogeneous mass. 
Replacement of the homogeneous mass $m_0$ with a mass $M\left( x \right)$ varying with position drastically complicates the harmonic oscillator problem. 
The concept of a position-dependent mass can be traced up to the seminal Giaever experiment on the observation of the tunneling effect in superconductor structures~\cite{giaever1960a,giaever1960b}, where the band structure varying with position is introduced for a complete theoretical explanation of the observed phenomena~\cite{harrison1961}, and the concept of the position-dependent band structure later has been generalized to the idea of the position-dependent mass~\cite{bendaniel1966}. 
Since its introduction, the concept of the mass varying with position has found successful applications in various branches of physics and related areas. 
Its success is evident from the large amount of publications in scientific journals, devoted to its applications~\cite{zhu1983,vonroos1983,morrow1984,li1993,plastino1999,kolesnikov1999,li2009,lima2012,nobre2015,morris2015,amir2016,karthiga2017,mustafa2019,elnabulsi2020}. 
The very sensitive point of these published works is that all of them are done within the canonical commutation relations. 
We decided to revisit this problem from a different angle -- to introduce an analytical expression of the position-dependent mass to the quantum harmonic oscillator in such a way that it will lead to a parameter deformation of the harmonic oscillator Heisenberg-Weyl algebra and then to generalize this case to the more complicated Lie superalgebra $\mathfrak{osp}\left( {1|2} \right)$. 

Our motivation for revisiting the problem from the physics point of view is based on the goal of constructing an exactly solvable oscillator model that will generalize a triangular-shaped potential on the one hand and an infinite potential well with a non-rectangular profile on the other hand. 
Both these potentials are significant due to their wide use in experimental physics. 
As we are aware, the first detailed description of the triangular potential appears in~\cite{nagamiya1940a,nagamiya1940b}. 
Then, it was intensively applied as an intermolecular potential for a more accurate prediction of the peculiar behaviour and specific phase transitions of several many-particle systems~\cite{feinberg1964,card1972,benavides2007,benavides2018}. 
Later, interesting applications of the potentials with similar profiles also appeared in the field of low-dimensional systems, where they have been used for the calculation of the electronic states of the {\em InGaAs/GaAs} quantum wells of non-rectangular profile~\cite{vlaev1994}.
At the same time, recent achievements in the design of the one-dimensional photonic crystal cavities show that, unlike the traditional method of variation of the lattice constant, the cavity for the photons also can be successfully implemented via a variation of the crystal width~\cite{ahn2010,dobbelaar2015,massaro2024}. 
Then, even high-order modes in such photonic crystal cavities with varying widths exhibit Hermite-Gauss distribution with the equidistant decreasing of their spacing. 
It is well known that an exact solution to the infinite potential well problem leads to a non-equidistant energy spectrum. 
Moreover, one can observe equidistance of the energy levels in the models of the confined quantum systems with the oscillator-shaped profile, if there is an application of semiconfinement effect only~\cite{jafarov2021,jafarov2022,jafarov2023}. 
Such confined quantum systems are more appropriate for the successful description of the one-dimensional photonic crystal cavities with the varied lattice constant rather than the width of the crystal~\cite{bazin2014,marty2019}. 
In this case, we revised our computation technique and decided to construct a new model of the harmonic oscillator that in special cases can give birth to these specific potentials with an equidistant energy spectrum -- triangular-shaped potential and non-rectangular potential well.

We structured our paper as follows: a new deformation of the Heisenberg-Weyl algebra with a parameter $a>-1$ is introduced in Section 2 in the form of Proposition. 
The proof of the Proposition is presented, too. 
Section 3 is devoted to the analytical realization of the corresponding non-relativistic quantum harmonic oscillator within the canonical approach, the dynamical symmetry algebra of which is this deformed Heisenberg-Weyl algebra. 
Next, the deformed Heisenberg-Weyl algebra introduced in Section 2 is generalized to the deformed $\mathfrak{osp}\left( {1|2} \right)$ Lie superalgebra in Section 4. 
Such a deformation is presented in the form of another proposition and its proof is provided, too. 
In Section 5, we present an analytical realization of the corresponding non-relativistic quantum harmonic oscillator within the non-canonical approach, the dynamical symmetry algebra of which is a deformed $\mathfrak{osp}\left( {1|2} \right)$ Lie superalgebra. 
The final section contains comparative plots of the computed energy spectrum versus the generalized oscillator potential and briefly discusses the obtained results.

\section{Deformation of the quantum harmonic oscillator Heisenberg-Weyl algebra with a parameter $a>-1$}

As highlighted in the Introduction, our starting point is the following first-order differential operator realization of the momentum operator in the $x$-position representation:

\be
\label{p-nr}
\hat p_x  =  - i\hbar \frac{d}{{dx}}.
\ee
Its commutation with the position operator defined in the $x$-representation as

\be
\label{x-nr}
\hat x = x,
\ee
is well known and it is called the canonical commutation relation between the momentum and position operators:

\be
\label{ccr}
\left[ {\hat p_x ,\hat x} \right] = - i\hbar .
\ee
 
Introduction of the quantum harmonic oscillator creation and annihilation operators $\hat a^ + $ and $\hat a^ - $ defined through $\hat p_x$ and $\hat x$ as follows:

\be
\label{caop-can}
\begin{split}
\hat a^ +   = \frac{1}{{\sqrt 2 }}\left( {\sqrt {\frac{{m_0 \omega }}{\hbar }} \hat x - \frac{i}{{\sqrt {m_0 \omega \hbar } }}\hat p_x } \right),\\
\hat a^ -   = \frac{1}{{\sqrt 2 }}\left( {\sqrt {\frac{{m_0 \omega }}{\hbar }} \hat x + \frac{i}{{\sqrt {m_0 \omega \hbar } }}\hat p_x } \right),
\end{split}
\ee
leads to the Heisenberg-Weyl algebra of the quantum harmonic oscillator, involving the creation-annihilation operators $\hat a^ \pm$ and the Hamiltonian $\hat H$:

\be
\label{hw-alg}
\begin{split}
 \left[ {\hat H,\hat a^ \pm  } \right] &=  \pm \hbar \omega \hat a^ \pm  , \\ 
 \left[ {\hat a^ -  ,\hat a^ +  } \right] &= 1.
\end{split}
\ee
Here, the Hamiltonian with the quantum harmonic oscillator potential

\be
\label{osc-pot}
V\left(x\right) = \frac{{m_0 \omega ^2 \hat x \cdot \hat x}}{2},
\ee
is defined in the non-relativistic approach as

\be
\label{osc-h}
\hat H = \hbar \omega \left( {\hat a^ +  \hat a^ -   + \frac{1}{2}} \right) = \frac{{\hat p_x  \cdot \hat p_x }}{{2m_0 }} + \frac{{m_0 \omega ^2 \hat x \cdot \hat x}}{2},
\ee
with $m_0$ being the homogeneous mass of the oscillator and $\omega$ being as its angular frequency.

\begin{proposition}
\label{prop1}
Let $M \equiv M \left( x \right)$ be a position-dependent mass that is introduced for the non-relativistic harmonic oscillator system instead of its constant mass $m_0$. 
Then, the replacement 
\be
\label{subst1}
\sqrt{m_0} \hat x \to \sqrt{M(x)}\hat x, \qquad
\frac{1}{{\sqrt {m_0 } }}\hat p_x  \to \frac{1}{{M^{\frac{1}{4}} }}\hat p_x \frac{1}{{M^{\frac{1}{4}} }},
\ee
which preserves the hermiticity of the momentum operator in the $x$-configurational representation,
in the Hamiltonian $\hat H$ \eqref{osc-h} and in the operators $\hat a^\pm$ \eqref{caop-can}, 
leads to the following deformation of the Heisenberg-Weyl algebra with a parameter $a>-1$:
\be
\label{a-hw-alg}
\begin{split}
\left[ {\hat H,\hat a^ \pm  } \right] &=  \pm \hbar \omega \left( {1 + a} \right)\hat a^ \pm  ,\\
\left[ {\hat a^ -  ,\hat a^ +  } \right] &= 1 + a,
\end{split}
\ee
if the mass $M\left( x \right)$ is of the following form:
\be
\label{m-pd}
M \left( x \right) = m_0 \left( {\lambda _0^2 x^2} \right)^a = m_0 \left| {\lambda _0 x} \right|^{2a} , \qquad \lambda _0  = \sqrt {\frac{{m_0 \omega }}{\hbar }}.
\ee
\end{proposition}

\textbf{Proof}. 
First of all, note that the introduction of the combination of product of the position-dependent mass and momentum operator as $\frac{1}{{M^{\frac{1}{4}} }}\hat p_x \frac{1}{{M^{\frac{1}{4}} }}$ is not new. This point is discussed thoroughly in~\cite{mustafa2007}.

Proposition~\ref{prop1} is proven by the direct computations of the above commutation relations. 
First, one finds:

\be
\label{ccr-pdm}
\left[ {\frac{1}{{M^{\frac{1}{4}} }}\hat p_x \frac{1}{{M^{\frac{1}{4}} }},M^{\frac{1}{2}} \hat x} \right] =  - i\hbar \left( {1 + \frac{1}{2}\frac{{M'}}{M}} x \right),
\ee
where $M'\equiv dM/dx$.

Applying the substitutions \eqref{subst1} in \eqref{caop-can}, yields:

\be
\label{ccr-gen}
\left[ {\hat a^ -  ,\hat a^ +  } \right] = \frac{i}{\hbar }\left[ {\frac{1}{{M^{\frac{1}{4}} }}\hat p_x \frac{1}{{M^{\frac{1}{4}} }},M^{\frac{1}{2}} \hat x} \right] = 1 + \frac{1}{2}\frac{{M'}}{M} x.
\ee

In order to have a simple deformation of the Heisenberg-Weyl algebra by means of a parameter $a$ as in \eqref{a-hw-alg}, one needs

\be
\label{m-def}
\frac{1}{2}\frac{{M'}}{M}x = a, \quad a>-1.
\ee
From here, one easily computes the integral and obtains that $M = m_0 \left( {\lambda _0^2 x^2} \right)^a= m_0 \left| {\lambda _0 x} \right|^{2a}$. 
Here, $\lambda _0$ is introduced through the integration constant aiming to preserve the dimensionless nature of the multiplier of mass $m_0$. Its substitution in the commutation relations \eqref{ccr-pdm} and \eqref{ccr-gen} proves the second commutation relation of eq.\eqref{a-hw-alg}. The first relation of \eqref{a-hw-alg} is computed in an analogous way. 

\qedsymbol

At this point, the restriction $a>-1$ imposed in eq.~\eqref{m-def} seems artificial.
In fact, if one wants to avoid singularities for the function $M(x)$, one would rather just impose $a\geq 0$.
The essential justification for the condition $a>-1$ is given only in the following section: 
in order to have deformations of the harmonic oscillator wavefunctions with appropriate boundary conditions, 
it is necessary to avoid singularities for the harmonic oscillator potential at $x=0$.
The fact that $M(x)$ has a singularity at $x=0$ for $-1<a<0$ does not lead to a singularity of the potential, 
but for values $a<-1$ it would lead to such a singularity of the potential.

One also observes that the case $a=-1$ destroys the non-commuting nature of the algebra generators in \eqref{a-hw-alg}. 
The vanishing of this commutator would imply a violation of the uncertainty principle, which is a central concept of quantum mechanics.

The deformation of the Heisenberg-Weyl algebra \eqref{a-hw-alg} by means of a simple factor $(1+a)$ will turn out to be very relevant.
We are going to present the analytical realization for the corresponding non-relativistic quantum harmonic oscillator within the canonical approach.
The exact expressions of the wavefunctions of the stationary states and the energy spectrum of the discrete levels elegantly generalize those of the well-known oscillator.

\section{Non-relativistic harmonic oscillator with a deformed Heisenberg-Weyl dynamical symmetry algebra}

Our starting point is the following Schr\"odinger equation for the oscillator:

\be
\label{seq-01}
\left[ {\frac{{\hat{\tilde{p}}_x  \cdot \hat{\tilde{p}}_x }}{2} + \frac{{\omega ^2 }}{2}\hat{\tilde x} \cdot \hat{\tilde x}} \right]\psi \left( x \right) = E\psi \left( x \right),
\ee
where,
\be
\label{px-x}
\begin{split}
\hat{\tilde{p}}_x &= \frac{1}{{M^{\frac{1}{4}} }}\hat p_x \frac{1}{{M^{\frac{1}{4}} }} \equiv - i\hbar \frac{1}{{\lambda _0 ^a \sqrt {m_0 } }}\left( {\left| x \right|^{ - a} \frac{d}{{dx}} - \frac{a}{2}\left| x \right|^{ - a - 2}  \cdot x} \right), \\
\hat{\tilde x} &= M^{\frac{1}{2}} \hat x \equiv \sqrt {m_0 } \lambda _0 ^a \left| x \right|^{a } \cdot x .
\end{split}
\ee
Substitution of \eqref{px-x} in the Schr\"odinger equation \eqref{seq-01} leads to the following second order differential equation:

\be
\label{seq-02}
\left[ { - \frac{{\hbar ^2 }}{{2m_0 }}\lambda _0 ^{ - 2a} \left( {\left| x \right|^{ - a} \frac{d}{{dx}} - \frac{a}{2}\left| x \right|^{ - a - 2}  \cdot x} \right) \left( {\left| x \right|^{ - a} \frac{d}{{dx}} - \frac{a}{2}\left| x \right|^{ - a - 2}  \cdot x} \right) + \frac{{m_0 \omega ^2 }}{2}\lambda _0 ^{2a} \left| x \right|^{2a} \cdot x^2} \right]\psi \left( x \right) = E\psi \left( x \right).
\ee
Here, one observes that the harmonic oscillator potential would become singular at $x=0$ if $a<-1$.
This would lead to wavefunctions which are unbounded at $\pm \infty$.
The case $a=-1$ was already excluded in the previous section. 
Note that for $a=-1$ the potential is constant and also then there would be no bound state solutions in $-\infty<x<+\infty$.
All together, we shall from now onwards impose the restriction $a>-1$.

Multiplication of this equation by $\left(- \frac{{2m_0 }}{{\hbar ^2 }}\cdot \lambda _0 ^{2a} \right)$ slightly changes it as follows:

\be
\label{seq-03}
\left[ {\left( {\left| x \right|^{ - a} \frac{d}{{dx}} - \frac{a}{2}\left| x \right|^{ - a - 2}  \cdot x} \right)\left( {\left| x \right|^{ - a} \frac{d}{{dx}} - \frac{a}{2}\left| x \right|^{ - a - 2}  \cdot x} \right) + \frac{{2m_0 E}}{{\hbar ^2 }}\lambda _0 ^{2a}  - \lambda _0 ^{4a + 4} \left| x \right|^{2a} \cdot x^2 } \right]\psi \left( x \right) = 0.
\ee
Simple computations show that

\be
\label{seq-04}
\begin{split}
\left( {\left| x \right|^{ - a} \frac{d}{{dx}} - \frac{a}{2}\left| x \right|^{ - a - 2}  \cdot x} \right)&\left( {\left| x \right|^{ - a} \frac{d}{{dx}} - \frac{a}{2}\left| x \right|^{ - a - 2}  \cdot x} \right) \\
&=  \left| x \right|^{ - 2a} \left[\frac{{d^2 }}{{dx^2 }} - \frac{2a}{x} \frac{d}{{dx}} +  \frac{\frac{1}{4}a\left( {3a + 2} \right)}{x^2}\right] .
\end{split}
\ee
Using this in \eqref{seq-03} and further multiplying the equation by $\left| x \right|^{2a}$ leads to

\be
\label{seq-05}
\psi '' - \frac{{2a}}{x}\psi ' + \left[ {\frac{1}{4}a\left( {3a + 2} \right)x^{ - 2}  + \kappa ^2 \lambda _0 ^{2a} \left| x \right|^{2a}  - \lambda _0 ^{4a + 4} \left| x \right|^{4a}\cdot x^2 } \right]\psi  = 0.
\ee
Here, the notations $\psi \equiv \psi \left( x \right)$, $\frac{{d^2 \psi }}{{dx^2 }} \equiv \psi ''$, $\frac{{d \psi }}{{dx }} \equiv \psi '$ and $\kappa  = \sqrt {\frac{{2m_0 E}}{{\hbar ^2 }}}$ are applied in the above equation for simplicity.

Now, it is convenient to introduce a dimensionless variable

\be
\label{xi-x}
\xi  = \frac{{\lambda _0 ^{a + 1} }}{{\sqrt {a + 1} }}\left| x \right|^a \cdot x, \qquad -\infty < \xi < +\infty .
\ee
Such a transformation from $x$ to $\xi$ carried out through eq.~\eqref{xi-x} is a well-known point canonical transformation~\cite{manning1935,bhattacharjie1962,pak1984}. 
It can be successfully applied to the exact solution of the second-order differential equation for the different forms of quantum systems with position-dependent mass~\cite{alhaidari2002,alhaidari2003,quesne2004,gang2005,mustafa2006,tezcan2007,quesne2009,quesne2016}. 
Its substitution in eq.~\eqref{seq-05}, and multiplication by $\left( {a + 1} \right)^{-\frac{{2a + 1}}{{a + 1}}} \cdot \lambda _0 ^{-2} \cdot \xi ^{-\frac{{2a}}{{a + 1}}}$, yields

\be
\label{seq-06}
\psi '' - \frac{a}{{\left( {a + 1} \right)\xi }}\psi ' + \frac{{\frac{1}{4}a\left( {3a + 2} \right) + \left( {a + 1} \right)\kappa ^2 \lambda _0 ^{ - 2} \xi ^2  - \left( {a + 1} \right)^2 \xi ^4 }}{{\left( {a + 1} \right)^2 \xi ^2 }}\psi  = 0.
\ee

We look for analytical solutions of the following form:

\be
\label{psi-xi}
\psi  = f\left(\xi\right) \cdot y\left(\xi\right),
\ee
where $f\left(\xi\right)$ defines the boundary behavior of the wavefunction of the stationary states, and $y\left(\xi\right)$ determines the polynomial behavior of the quantum system within the harmonic oscillator potential. 
In particular, since we are interested in deformations of harmonic oscillator wavefunctions only, we will investigate analytical expression of $f\left(\xi\right)$ with parameters $A$ and $B$ as

\be
\label{f-xi}
f = \left| \xi \right| ^A e^{B\xi ^2 } .
\ee
Herein, the parameter $A$ is responsible for the continuity property of the wavefunction $\psi$ at $x=0$, whereas the parameter $B$ (which should be negative) is responsible for its vanishing at $x=\pm \infty$.

Further easy computations now lead to the following second order differential equation for $y$:

\be
\label{seq-07}
\begin{split}
&y'' + \frac{{2A - \frac{a}{{a + 1}} + 4B\xi ^2 }}{\xi }y' + \\
&\Scale[0.99]{ + \frac{{\left( {a + 1} \right)^2 A^2  - \left( {a + 1} \right)\left( {2a + 1} \right)A + a/2\left( {3a/2 + 1} \right) + \left[ {2B\left( {2A + 1} \right)\left( {a + 1} \right) - 2aB + \kappa ^2 \lambda _0 ^{ - 2} } \right]\left( {a + 1} \right)\xi ^2  + \left( {4B^2  - 1} \right)\left( {a + 1} \right)^2 \xi ^4 }}{{\left( {a + 1} \right)^2 \xi ^2 }}y = 0.}
\end{split}
\ee
One observes that the above equation has indeed a polynomial solution of the function $y\left(\xi\right)$ only if the following condition is satisfied:

\be
\label{mu-01}
\begin{split}
& \left( {a + 1} \right)^2 A^2  - \left( {a + 1} \right)\left( {2a + 1} \right)A + a/2\left( {3a/2 + 1} \right) + \\
& \left[ {2B\left( {2A + 1} \right)\left( {a + 1} \right) - 2aB + \kappa ^2 \lambda _0 ^{ - 2} } \right]\left( {a + 1} \right)\xi ^2  + \left( {4B^2  - 1} \right)\left( {a + 1} \right)^2 \xi ^4  = \mu \left( {a + 1} \right)^2 \xi ^2 ,
\end{split}
\ee
where the parameter $\mu$ will be determined later on.

Equation \eqref{mu-01} is a polynomial in $\xi$ which should be identically zero. 
This leads immediately to the solutions for $A$ and $B$. 
For $B$, the solutions are $1/2$ and $-1/2$; but the boundary conditions for the wavefunctions imply that only $B=-1/2$ can be retained.
For $A$, the quadratic equation following from \eqref{mu-01} leads to

\be
\label{a-eps}
A = 1 + \frac{1}{2}\left( {\varepsilon  - \frac{1}{{a + 1}}} \right),
\ee
where $\varepsilon=\pm1$. Then 

\be
\label{mu-eps}
\mu  = \left( {\varepsilon + 2} \right) + \frac{{\kappa ^2 \lambda _0 ^{ - 2} }}{{a + 1}}.
\ee
The substitution of \eqref{a-eps} and $B=-1/2$ in \eqref{seq-07} leads to 

\be
\label{seq-09}
y'' + \left[ {\left( {\varepsilon  + 1} \right)\xi ^{ - 1}  - 2\xi } \right]y' + \left( {\frac{{\kappa ^2 \lambda _0 ^{ - 2} }}{{a + 1}} - \varepsilon  - 2} \right)y = 0.
\ee
If $\varepsilon  =  + 1$ one obtains the following value of $A$

\be
\label{a1-eps}
A_1  = 1 + \frac{1}{2}\frac{a}{{a + 1}},
\ee
whereas for $\varepsilon  =  - 1$ one obtains
\be
\label{a2-eps}
A_2  = \frac{1}{2}\frac{a}{{a + 1}}.
\ee
One observes that $A_1=A_2+1$, which means that the corresponding solutions of \eqref{seq-09} (denoted by $y_1$ and $y_2$ respectively) satisfy $y_2= \xi y_1$. 
Hence, it is sufficient to continue with $A_2$ (and $\varepsilon  =  - 1$) only.
Equation~\eqref{seq-09} simplifies then as follows:

\be
\label{seq-11}
y_2'' - 2\xi y_2' + \frac{{\kappa ^2 \lambda _0 ^{ - 2}  - a - 1}}{{a + 1}}y_2 = 0.
\ee
The function $f$ reduces to
\be
\label{f-xi3}
f = \left| \xi \right| ^{\frac{a}{{2\left( {a + 1} \right)}}} e^{ - \frac{1}{2}\xi ^2 } .
\ee
The solutions of \eqref{seq-11} for $y_2$ are Hermite polynomials of degree $n$:

\be
\label{wf-y}
y_2  = H_n \left( \xi  \right), \qquad n=0,1,2,\ldots.
\ee

The energy spectrum $E_n$ can be easily found from the relation

\be
\label{kappa-en}
\kappa ^2 \lambda _0 ^{ - 2}  - a - 1 = 2n\left( {a + 1} \right).
\ee
Easy computations yield

\be
\label{e-kappa}
E_n  = \left( {a + 1} \right)\hbar \omega \left( {n + \frac{1}{2}} \right), \qquad n=0,1,2,\dots .
\ee
Our wavefunction has the following analytical expression:

\be
\label{psi-fin}
\psi _n \left( x \right) = C_n \sqrt[4]{{\left( {\lambda _0 ^2 x^2 } \right)^a }}e^{ - \frac{{\left( {\lambda _0^2 x^2} \right)^{a + 1} }}{{2\left( {a + 1} \right)}}} H_n \left( {\frac{{\lambda _0 x}}{{\sqrt {a + 1} }}\sqrt {\left( {\lambda _0 ^2 x^2 } \right)^a } } \right) .
\ee
Note the form of the exponential function in this expression: the vanishing at $\pm\infty$ requires indeed $a>-1$, as mentioned earlier.

The normalization coefficient $C_n$ can be computed using the orthogonality relation for Hermite polynomials~\cite{koekoek2010}:

\[
\frac{1}{{\sqrt \pi  }}\int\limits_{ - \infty }^{ + \infty } {e^{ - x^2 } H_m \left( x \right)H_n \left( x \right)dx}  = 2^n n!\delta _{mn} .
\]
Easy computations yield:

\be
\label{cn-hw}
C_n  = \left(\frac{a+1}{{\pi  }}\right)^{\frac{1}{4}} \sqrt {\frac{{\lambda _0  }}{{2^n n!}}} .
\ee

For the quantum oscillator under consideration, we managed to give closed form expressions for the wavefunctions of the stationary states~\eqref{psi-fin}, and for the discrete energy spectrum~\eqref{e-kappa}.
We are going to generalize these results to the case of the parabose oscillator in the next two sections.
Then all obtained results will be discussed briefly and their possible limit cases will be analyzed in the final section.

\section{Deformation of the quantum parabose oscillator algebra}

Now we generalize our analysis presented in Section 2 to the case where our starting point is the following first-order differential operator realization of the momentum operator in the $x$-position representation:

\be
\label{p-pb}
\hat p_x  =  - i\hbar \left( {\frac{d}{{dx}} - \frac{{\gamma  - 1/2}}{x}\hat R} \right), \qquad \gamma>1/2.
\ee
Here, $\hat R$ is a reflection operator~\cite{plyushchay1997}: $\hat R f(x)=f(-x)$.
Definition \eqref{p-pb} easily recovers the canonical definition \eqref{p-nr} if $\gamma=1/2$. 
Its commutation with the position operator \eqref{x-nr} is called a non-canonical commutation relation and it is defined in the position $x$-representation as

\be
\label{nccr}
\left[ {\hat p_x ,\hat x} \right] = - i\hbar \left[ {1 + \left( {2\gamma  - 1} \right)\hat R} \right].
\ee
Introduction of the quantum harmonic oscillator creation and annihilation operators $\hat a^ +  $ and $\hat a^ - $ as in \eqref{caop-can}, but with $\hat p_x$ given by \eqref{p-pb}, leads to the following relation:
\be
\label{parabose}
 \left[ {\hat a^ -  ,\hat a^ +  } \right] = 1 + \left( {2\gamma  - 1} \right)\hat R.
\ee
Together with $\{ \hat R, \hat a^\pm\}=0$, this forms the so-called parabose oscillator algebra~\cite{plyushchay1997}.

The Hamiltonian of the quantum parabose oscillator is defined in the non-relativistic approach in a slightly different manner than \eqref{osc-h}. It is as follows:

\be
\label{pbosc-h}
\hat H = \frac{\hbar \omega}2 \left( {\hat a^ +  \hat a^ -   + \hat a^ -  \hat a^ +} \right) = \frac{{\hat p_x  \cdot \hat p_x }}{{2m_0 }} + \frac{{m_0 \omega ^2 \hat x \cdot \hat x}}{2}.
\ee

The algebra generated by $\hat H$ and $\hat a^\pm$ (as odd generators) is the Lie superalgebra $\mathfrak{osp}\left( {1|2} \right)$.
Note that

\be
\label{osp-alg}
 \left[ {\hat H,\hat a^ \pm  } \right] =  \pm \hbar \omega \hat a^ \pm.
\ee

\begin{proposition}
\label{prop2}

Let $M \equiv M \left( x \right)$ be a position-dependent mass that is introduced for the non-relativistic harmonic oscillator system instead of its constant mass $m_0$. 
Then, the replacement \eqref{subst1}
in the Hamiltonian $\hat H$ \eqref{pbosc-h} and in the operators $\hat a^\pm$ \eqref{caop-can} 
(where $\hat p_x$ is as in \eqref{p-pb}), 
leads to the following deformation of the parabose algebra and of the Lie superalgebra $\mathfrak{osp}\left( {1|2} \right)$ with a parameter $a>-1$:
\be
\label{a-osp-alg}
\begin{split}
\left[ {\hat H,\hat a^ \pm  } \right] &=  \pm \hbar \omega \left( {1 + a} \right)\hat a^ \pm  ,\\
\left[ {\hat a^ -  ,\hat a^ +  } \right] &= 1 + a + \left( {2\gamma  - 1} \right)\hat R,
\end{split}
\ee
if the mass $M\left( x \right)$ is of the form \eqref{m-pd}.
\end{proposition}

\textbf{Proof}. 
As in Proposition \ref{prop1}, the proof is by direct computations of the above commutation relations. 
First, one computes the commutation between $\frac{1}{{M^{\frac{1}{4}} }}\hat p_x \frac{1}{{M^{\frac{1}{4}} }}$ and $M^{\frac{1}{2}} \hat x$. However, one needs to take into account that $\hat p_x$ now is defined through \eqref{p-pb}. The commutation relation yields:

\be
\label{nccr-pdm}
\left[ {\frac{1}{{M^{\frac{1}{4}} }}\hat p_x \frac{1}{{M^{\frac{1}{4}} }},M^{\frac{1}{2}} \hat x} \right] =  - i\hbar \left[ {1 + \frac{1}{2}\frac{{M_ +  '}}{{M_ +  }}x + \left( {\frac{{M_ +  ^{\frac{1}{4}} }}{{M_ -  ^{\frac{1}{4}} }} + \frac{{M_ -  ^{\frac{1}{4}} }}{{M_ +  ^{\frac{1}{4}} }}} \right)\left( {\gamma  - 1/2} \right)\hat R} \right].
\ee
Here, the additional definitions $M_ +   \equiv M\left( x \right)$ and $M_ -   \equiv M\left( - x \right)$ are introduced for convenience.

We consider the following case very similar to eq. \eqref{m-def}:

\be
\label{m-pb-def}
\frac{1}{2}\frac{{M_ +  '}}{{M_ +  }}x \equiv \frac{1}{2}\frac{{M'}}{M}x = a, \quad a>-1.
\ee
Then, we have

\be
\label{m+}
M_ +   = m_0 \left( {\lambda _0 ^2 x^2 } \right)^a = m_0 \left| {\lambda _0 x} \right|^{2a} ,
\ee
and since this is an even function:

\be
\label{m-}
M_ -   = M_ +  .
\ee
This means that the following additional conditions for the action of the reflection operator $\hat R$ on the position-dependent mass $M\left( x \right)$ hold:

\be
\label{r-mx}
\begin{split}
\hat RM^\nu  \left( x \right)& = M^\nu  \left( x \right), \\ 
\hat RM' \left( x \right) &= M' \left( { - x} \right), \\ 
\left[ {\hat R,M^\nu  } \right] &= \hat RM^\nu   - M^\nu  \hat R = 0.
\end{split}
\ee

Next, the substitution of \eqref{m+}\&\eqref{m-} in \eqref{nccr-pdm} leads to

\be
\label{nccr-pdm2}
\left[ {\frac{1}{{M^{\frac{1}{4}} }}\hat p_x \frac{1}{{M^{\frac{1}{4}} }},M^{\frac{1}{2}} \hat x} \right] =  - i\hbar \left[ {1 + a + \left( {2\gamma  - 1} \right)\hat R} \right].
\ee
This commutation relation further results in the correctness of all three commutation relations from \eqref{a-osp-alg}. 
We drop these computations here because they are tedious but straightforward to perform.

\qedsymbol

As before, the condition $a>-1$ is relevant only when the actual wavefunctions of the system are considered, but we impose it already now.

The concept of the position-dependent mass already led to an interesting deformation of the Heisenberg-Weyl algebra~\eqref{a-hw-alg}.
The introduction of this concept to the parabose oscillator has more far-reaching consequences.
Now the parity operator $\hat R$ appears as a result of the commutation relation between the parabose oscillator creation $a^ +$ and annihilation $a^ -$ operators.
This deformed superalgebra defined through~\eqref{a-osp-alg} reduces to the deformed Heisenberg-Weyl algebra~\eqref{a-hw-alg} if $\gamma=1/2$ or it returns to the known Lie superalgebra $\mathfrak{osp}\left( {1|2} \right)$ if $a=0$. 
Therefore, the realization of the parabose quantum oscillator model possessing such a dynamical symmetry algebra in terms of the exact analytical expressions of the energy spectrum and wavefunctions of the stationary states becomes very attractive. 
In the next section, we are going to show how such an oscillator model with deformed parabose statistics can be constructed.

\section{Non-relativistic parabose oscillator with a deformed $\mathfrak{osp}\left( {1|2} \right)$ dynamical symmetry superalgebra}

Our starting point is the Schr\"odinger equation for the parabose oscillator completely overlapping with eq. \eqref{seq-01}, 
but now with the operator $\hat{\tilde{p}}_x$ defined as
\be
\label{px-osp}
\hat{\tilde{p}}_x = \frac{1}{{M^{\frac{1}{4}} }}\hat p_x \frac{1}{{M^{\frac{1}{4}} }} \equiv - i\hbar \frac{1}{{M^{\frac{1}{2}} }}\left( {\frac{d}{{dx}} - \frac{1}{4}\frac{{M'}}{M} - \frac{{\gamma  - 1/2}}{x}\hat R} \right), 
\ee
whereas the operator $\hat{\tilde x}$ preserves its expression as defined in \eqref{px-x}.

Substitution of \eqref{px-osp} in the Schr\"odinger equation \eqref{seq-01} now leads to the following second-order differential equation:

\be
\label{oseq-02}
\begin{split}
 - \frac{{\hbar ^2 }}{{2m_0 }}&\lambda _0 ^{ - 2a} \left|x\right|^{ - 2a} \left( {\frac{{d^2 }}{{dx^2 }} - \frac{{2a}}{x}\frac{d}{{dx}} + \frac{{a\left( {3a + 2} \right)/4 - \left( {\gamma  - 1/2} \right)^2 +\left( {a + 1} \right)\left( {\gamma  - 1/2} \right) \hat R}}{{x^2 }}} \right) \psi \left( x \right) \\
&+ \frac{{m_0 \omega ^2 }}{2}\lambda _0 ^{2a} \left| x \right| ^{2a} \cdot x^2  \psi \left( x \right) = E\psi \left( x \right).
\end{split}
\ee
As in Section~3, multiplication with $\left(- \frac{{2m_0 }}{{\hbar ^2 }}\cdot \lambda _0 ^{2a} \right)$ leads to:

\be
\label{oseq-03}
\begin{split}
\left| x \right|^{ - 2a} &\left( \frac{{d^2 }}{{dx^2 }} - \frac{{2a}}{x}\frac{d}{{dx}} + \frac{{a\left( {3a + 2} \right)/4 - \left( {\gamma  - 1/2} \right)^2 +{\left( {a + 1} \right)\left( {\gamma  - 1/2} \right)}\hat R }}{{x^2 }} \right) \psi \left( x \right)\\
&+ \left(\kappa ^2 \lambda _0 ^{2a}  - \lambda _0 ^{4a + 4} \left| x \right|^{2a} \cdot x^2  \right)\psi \left( x \right) = 0.
\end{split}
\ee
Again, we introduced here $\kappa  = \sqrt {\frac{{2m_0 E}}{{\hbar ^2 }}} $.

Further, we follow the method applied for the construction of the undeformed parabose oscillator model in exact solutions~\cite{mukunda1980,ohnuki1982}:
the appearance of the operator $\hat R$ forces us to split the problem by studying even and odd solutions separately. 
First, we consider the even case when $n=2m$, where we write $\psi  \to \psi _{2m} $. 
Then, we have

\be
\label{oseq-05}
\begin{split}
\left[ {\left| x \right|^{ - 2a} \left( {\frac{{d^2 }}{{dx^2 }} - \frac{{2a}}{x}\frac{d}{{dx}} + \frac{{a\left( {3a + 2} \right)/4 - \left( {\gamma  - 1/2} \right)^2 +\left( {a + 1} \right)\left( {\gamma  - 1/2} \right)}}{{x^2 }} } \right) + \kappa ^2 \lambda _0 ^{2a}  - \lambda _0 ^{4a + 4} \left| x \right|^{2a} \cdot x^2 } \right]\psi _{2m} = 0.
\end{split}
\ee
Now, multiplication by $\left| x \right|^{2a}$ yields

\be
\label{oseq-06}
\begin{split}
&\psi _{2m} '' - 2ax^{ - 1} \psi _{2m} ' + \\
& + \left\{ \left[ {a\left( {3a + 2} \right)/4 + \left( a+1 \right)\left( {\gamma  - 1/2} \right) - \left( {\gamma  - 1/2} \right)^2 } \right]x^{ - 2}  + \kappa ^2 \lambda _0 ^{2a} \left| x \right|^{2a}  - \lambda _0 ^{4a + 4} \left| x \right|^{4a} \cdot x^2  \right\}\psi _{2m}  = 0.
\end{split}
\ee
Again, we introduced the notations $\frac{{d^2 \psi }}{{dx^2 }} = \psi ''$ and $\frac{{d\psi }}{{dx}} = \psi '$.

It is convenient to follow the method employed in Section 3 and to work with the dimensionless variable \eqref{xi-x}. 
Its substitution in eq.\eqref{oseq-06} as well as further multiplication by $\left( {a + 1} \right)^{-\frac{{2a + 1}}{{a + 1}}} \cdot \lambda _0 ^{-2} \cdot \xi ^{-\frac{{2a}}{{a + 1}}}$ yields

\be
\label{oseq-07}
\begin{split}
&\psi _{2m} '' - a\left( {a + 1} \right)^{ - 1} \xi ^{ - 1} \psi _{2m} ' + \\
& + \left\{ \left[ {a\left( {3a + 2} \right)/4 + \left( {a + 1} \right)\left( {\gamma  - 1/2} \right) - \left( {\gamma  - 1/2} \right)^2 } \right]\left( {a + 1} \right)^{ - 2} \xi ^{ - 2}  + \right. \\
&\left. +\kappa ^2 \left( {a + 1} \right)^{ - 1} \lambda _0 ^{ - 2}  - \xi ^2  \right\}\psi _{2m}  = 0.
\end{split}
\ee

It is clear that this equation generalizes eq.\eqref{seq-06} and recovers it if $\gamma=1/2$. 
Therefore, we follow the same technique that was applied for obtaining an exact solution of eq.\eqref{seq-06} and look for solutions of eq.\eqref{oseq-07} as in \eqref{psi-xi} with only a slight notational difference (namely $y \to y_{2m}$).

We drop long computations and note that the value of $B$ is again $-1/2$, whereas the value of $A$ generalizes \eqref{a-eps} as follows:

\be
\label{a-pb-eps}
A = 1 + \frac{1}{2}\left( {\varepsilon  - \frac{{1 + \varepsilon \left( {2\gamma  - 1} \right)}}{{a + 1}}} \right),
\ee
with $\varepsilon=\pm 1$.
Substitution of this value of $A$ and $B=-1/2$ yields the following second order differential equation for $y\equiv y_{2m}$:

\be
\label{oseq-08}
y_{2m} ''  + \left( {\frac{{a + 1 + \varepsilon \left( {a - 2\gamma + 2} \right)}}{{a + 1}}\xi ^{ - 1}  - 2\xi } \right)y_{2m} ' + \frac{{\kappa ^2 \lambda _0 ^{ - 2}  - 2\left( {a + 1} \right) - \varepsilon \left( {a - 2\gamma  + 2} \right)}}{{a + 1}}y_{2m}  = 0.
\ee

This resembles the following well-known second-order differential equation, whose exact solution is the Laguerre polynomial~\cite{koekoek2010}:

\[
xy''\left( x \right) + \left( {\alpha  + 1 - x} \right)y'\left( x \right) + ny\left( x \right) = 0.
\]
Here,

\[
y\left( x \right) = L_n^{\left( \alpha  \right)} \left( x \right).
\]
Introducing a new variable

\[
x = z^2 ,
\]
and performing very simple computations, one obtains that:

\be
\label{lag-eq}
y''\left( {z^2 } \right) + \left[ {\left( {2\alpha  + 1} \right)z^{ - 1}  - 2z} \right]y'\left( {z^2 } \right) + 4ny\left( {z^2 } \right) = 0.
\ee

Comparison of eqs.\eqref{oseq-08} and \eqref{lag-eq} leads to

\be
\label{alpha-2m}
\alpha  \equiv \alpha _{2m}  = \frac{1}{2}\varepsilon \left( {1 - \frac{{ {2\gamma  - 1} }}{{a + 1}}} \right).
\ee
Then, the polynomial solution of eq.\eqref{oseq-08} can be written down in terms of the Laguerre polynomials as follows:

\be
\label{y-2m-lag}
y_{2m} \left( \xi  \right) = L_m^{\left( {\alpha _{2m} } \right)} \left( {\xi ^2 } \right).
\ee

The corresponding energy spectrum of the even states has the following expression:

\be
\label{e-2m-alpha}
E \equiv E_{2m}  = \left( {a + 1} \right)\hbar \omega \left( {2m + 1 + \alpha _{2m} } \right).
\ee

Further, taking into account that the energy spectrum of the even states \eqref{e-2m-alpha} should recover both \eqref{e-nc} if $a=0$ and \eqref{e-kappa} if $\gamma = 1/2$, one should restrict oneself to the case with $\varepsilon=-1$. Then, one obtains the following value of $A_2$ that generalizes \eqref{a2-eps}:

\be
\label{a2-pb-eps}
A_2 = \frac{1}{2}\frac{{a +  {2\gamma  - 1} }}{{a + 1}}.
\ee 
Additionally, substitution of $\varepsilon=-1$ in the expression of the parameter $\alpha _{2m}$ changes eq.\eqref{alpha-2m} in the following manner:

\be
\label{alpha-2m-2}
\alpha _{2m}  = \frac{1}{2} \left( {\frac{{ {2\gamma  - 1} }}{{a + 1}} - 1} \right).
\ee
Then, the energy spectrum of the even states \eqref{e-2m-alpha} also simplifies as follows:

\be
\label{e-2m-alpha2}
E_{2m}  = \left( {a + 1} \right)\hbar \omega \left( {2m + \frac 12 + \frac{{ {\gamma  - 1/2} }}{{a + 1}} } \right).
\ee

The wavefunction of the even stationary states generalizing even states of the wavefunction \eqref{psi-fin} for case $\gamma \geq 1/2$ can be written in its exact analytical expression as follows:

\be
\label{psi-pb-2m-fin}
\psi _{2m} \left( x \right) = C_{2m} \sqrt[4]{{\left( {\lambda _0 ^2 x^2 } \right)^{a +   {2\gamma  - 1} } }}e^{ - \frac{{\left( {\lambda _0^2 x^2} \right)^{a + 1} }}{{2\left( {a + 1} \right)}}} L_m^{\left( {\alpha _{2m} } \right)} \left( {\frac{{\left( {\lambda _0^2 x^2} \right)^{a + 1} }}{{a + 1}}} \right),
\ee
where the normalization factor $C_{2m}$ is equal to

\be
\label{c-pb-2m-fin}
C_{2m}  = \left( { - 1} \right)^m \left( {a + 1} \right)^{\frac{1}{2}\left( {\frac{1}{2} - \frac{{\gamma  - 1/2}}{{a + 1}}} \right)} \sqrt {\frac{{\lambda _0 m!}}{{\Gamma \left( {m + \alpha _{2m}  + 1} \right)}}} .
\ee
This can be easily computed from the orthogonality relation for the Laguerre polynomials \cite{koekoek2010}:

\be
\label{ort-lag}
\int\limits_0^\infty  {y^\alpha  e^{ - y} L_n^{\left( \alpha  \right)} \left( y \right)L_n^{\left( \alpha  \right)} \left( y \right)dy}  = \frac{{\Gamma \left( {n + \alpha  + 1} \right)}}{{n!}}.
\ee

One follows a similar procedure for the solution of the Schr\"odinger equation \eqref{oseq-03} for the odd states with $n=2m+1$, which reads
\be
\label{oseq-odd-01}
\begin{split}
\left[ {\left| x \right|^{ - 2a} \left( {\frac{{d^2 }}{{dx^2 }} - \frac{{2a}}{x}\frac{d}{{dx}} + \frac{{a\left( {3a + 2} \right)/4 - \left( {\gamma  - 1/2} \right)^2 -\left( {a + 1} \right)\left( {\gamma  - 1/2} \right)}}{{x^2 }} } \right) + \kappa ^2 \lambda _0 ^{2a}  - \lambda _0 ^{4a + 4} \left| x \right|^{2a} \cdot x^2 } \right]\psi _{2m+1} = 0.
\end{split}
\ee
Skipping these computations, we just give the main results.

The energy spectrum of the odd states is as follows:

\be
\label{e-2m1-alpha2}
E_{2m+1}  = \left( {a + 1} \right)\hbar \omega \left( {2m + \frac 32 + \frac{{ {\gamma  - 1/2} }}{{a + 1}} } \right).
\ee
The wavefunction of the odd stationary states generalizing odd states of the wavefunction \eqref{psi-fin} for the case $\gamma \geq 1/2$ can be written in its exact analytical expression as follows:

\be
\label{psi-pb-2m1-fin}
\psi _{2m+1} \left( x \right) = C_{2m+1} \sqrt[4]{{\left( {\lambda _0 ^2 x^2 } \right)^{3a +  {2\gamma  - 1} } }}e^{ - \frac{{\left( {\lambda _0^2 x^2} \right)^{a + 1} }}{{2\left( {a + 1} \right)}}} x L_m^{\left( {\alpha _{2m+1} } \right)} \left( {\frac{{\left( {\lambda _0^2 x^2} \right)^{a + 1} }}{{a + 1}}} \right),
\ee
where, the normalization factor $C_{2m+1}$ is equal to

\be
\label{c-pb-2m1-fin}
C_{2m+1}  = \left( { - 1} \right)^m \left( {a + 1} \right)^{ -\frac{1}{2}\left( {\frac{1}{2} + \frac{{\gamma  - 1/2}}{{a + 1}}} \right)} \sqrt {\frac{{\lambda _0 ^3 m!}}{{\Gamma \left( {m + \alpha _{2m + 1}  + 1} \right)}}} .
\ee

We have achieved our main goal of obtaining exact solutions to the harmonic oscillator problem having a dynamical symmetry defined by deformed parabose algebra, where the deformation parameter satisfies $a>-1$. 
In the final section, some basic properties of both canonical and parabose oscillator models with a position-dependent mass obtained in the present paper within the deformation of both the Heisenberg-Weyl algebra and the parabose algebra generalization are discussed.

\section{Discussions}

\begin{figure}[t!]
\begin{center}
\resizebox{0.32\textwidth}{!}{%
  \includegraphics{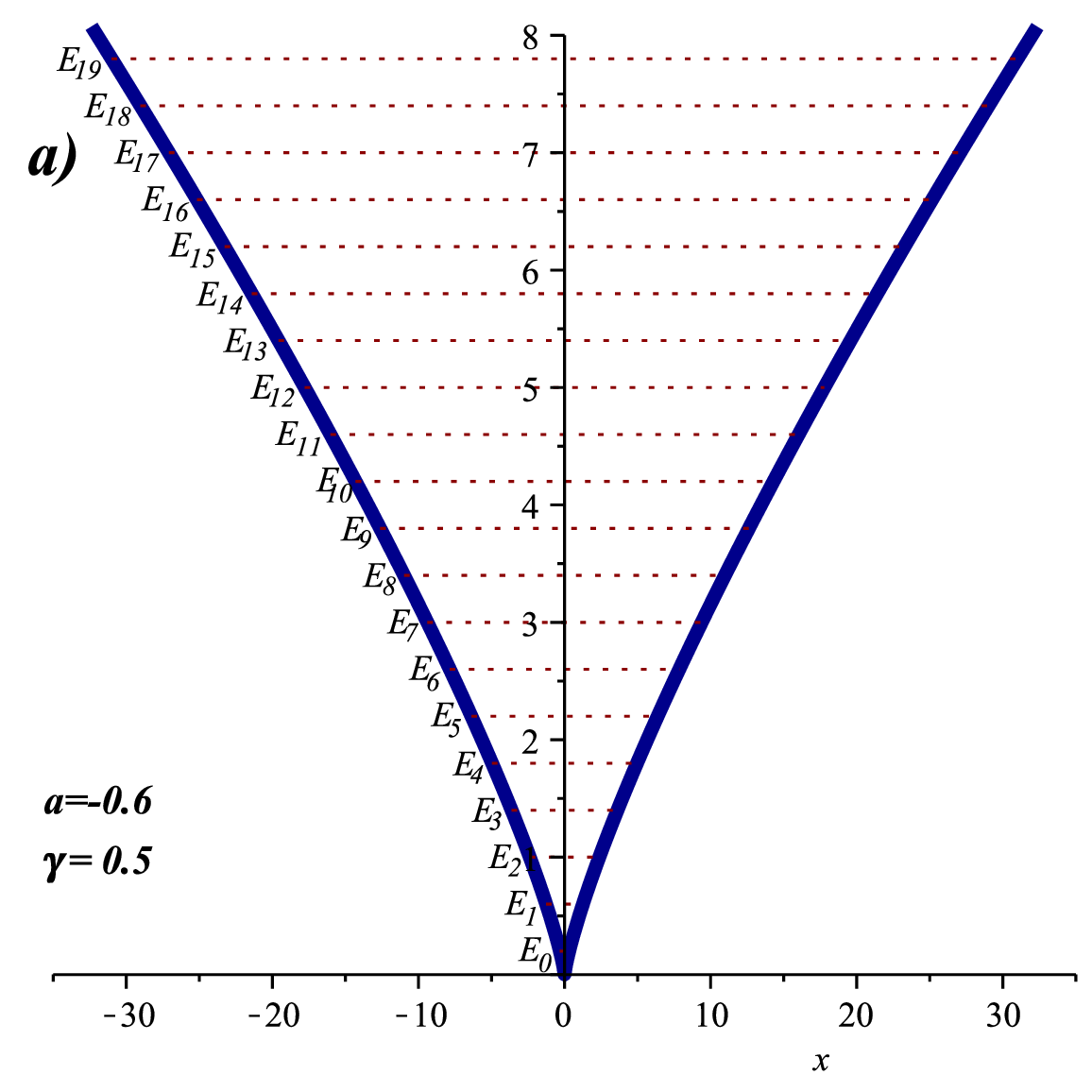}
}
\resizebox{0.32\textwidth}{!}{%
  \includegraphics{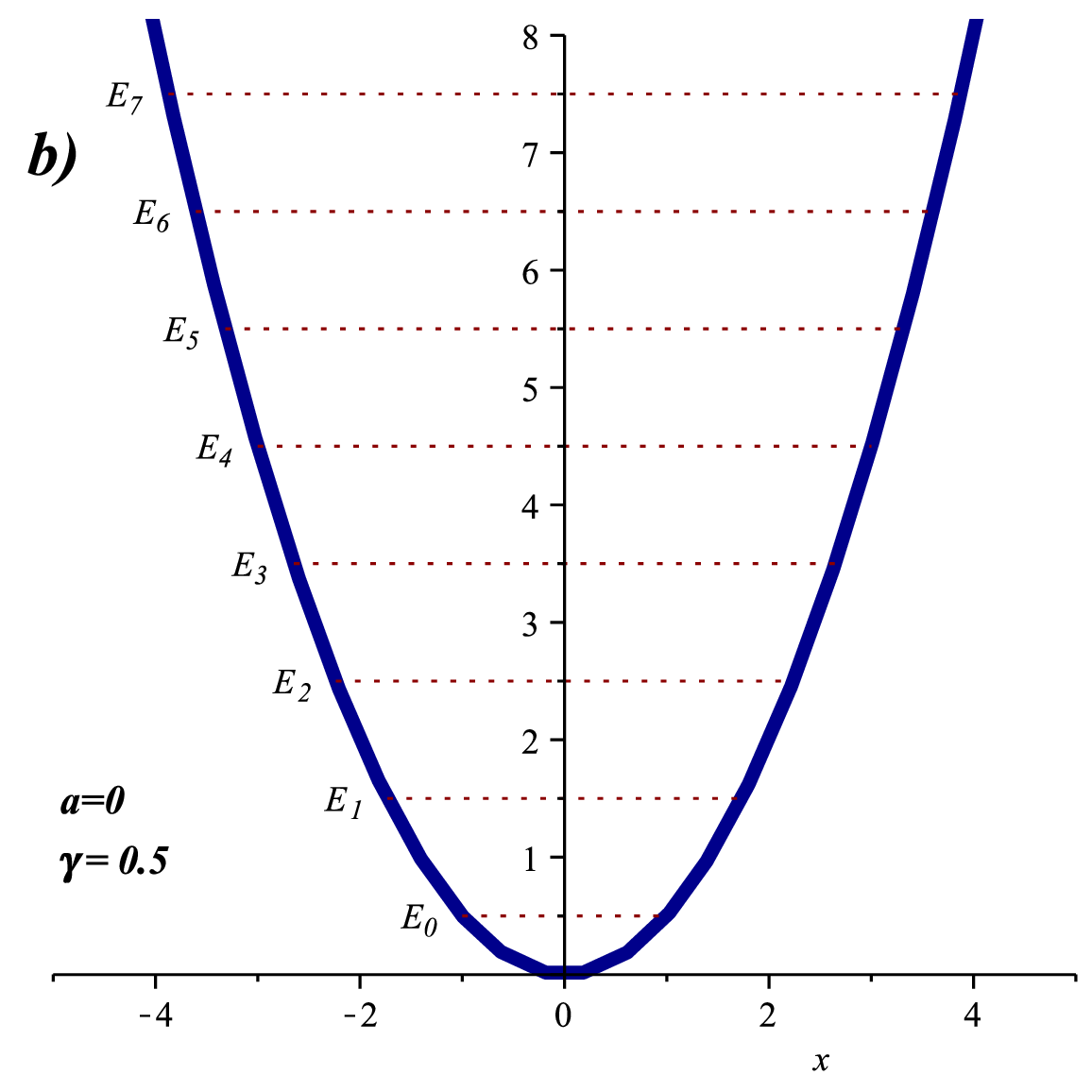}
}
\resizebox{0.32\textwidth}{!}{%
  \includegraphics{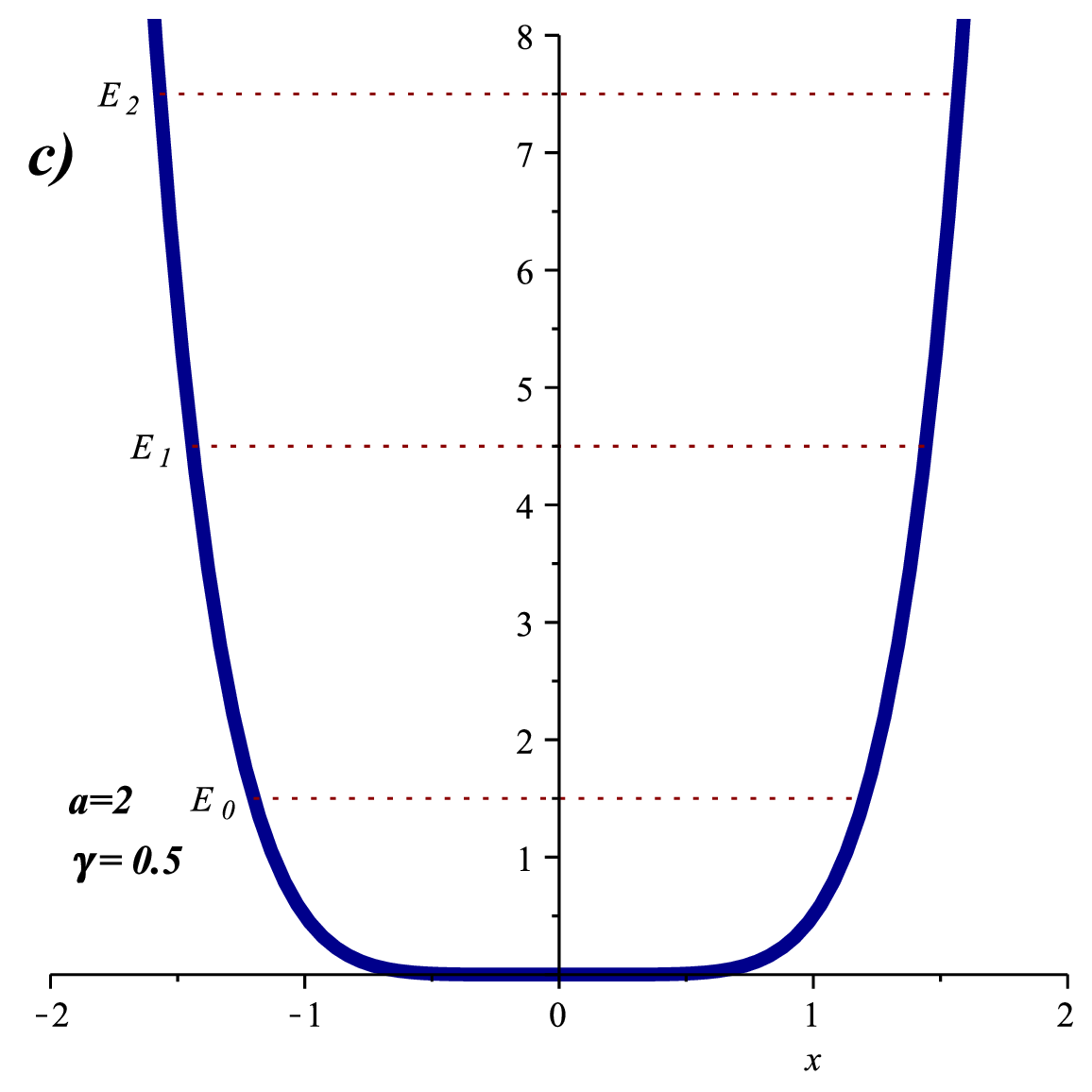}
}\\
\resizebox{0.32\textwidth}{!}{%
  \includegraphics{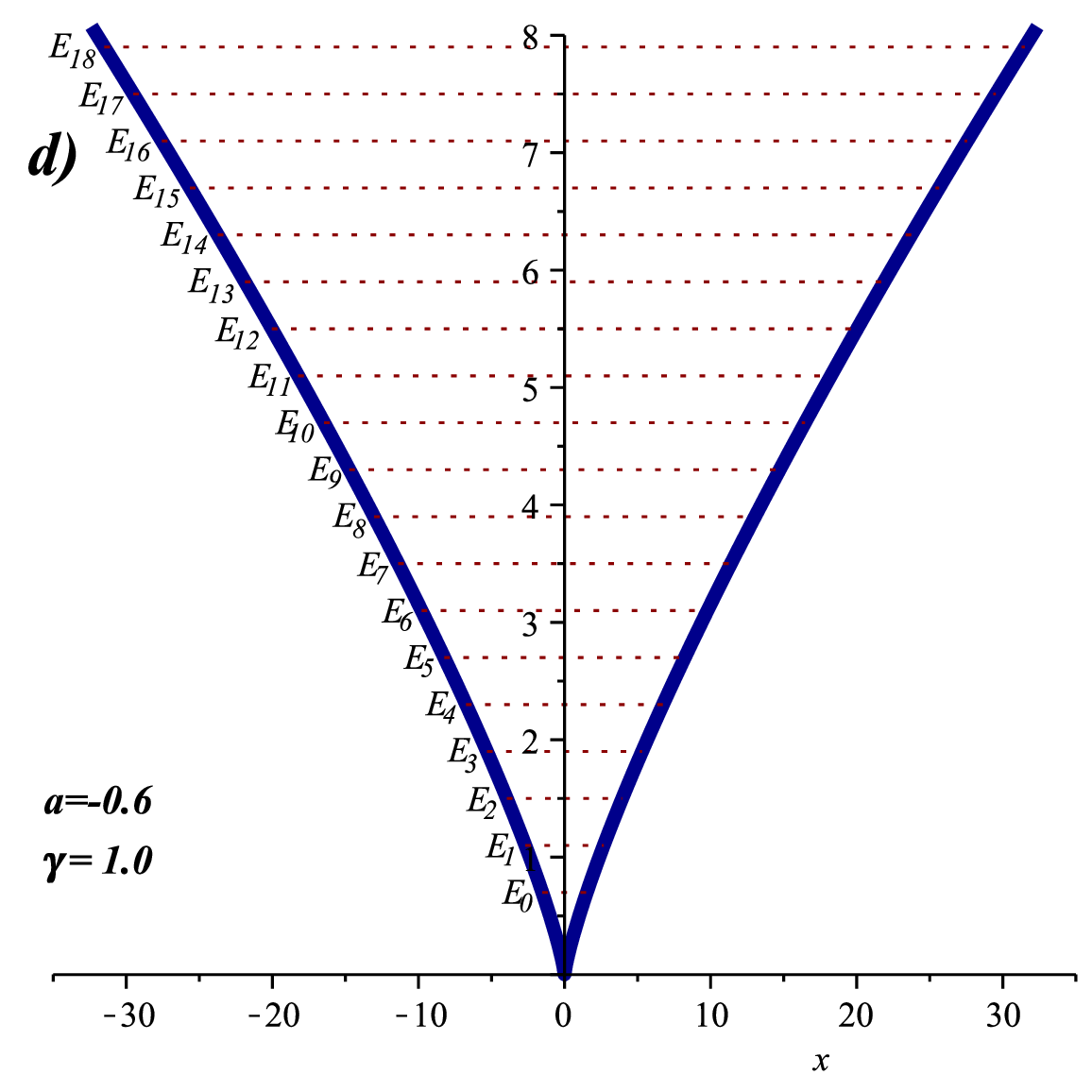}
}
\resizebox{0.32\textwidth}{!}{%
  \includegraphics{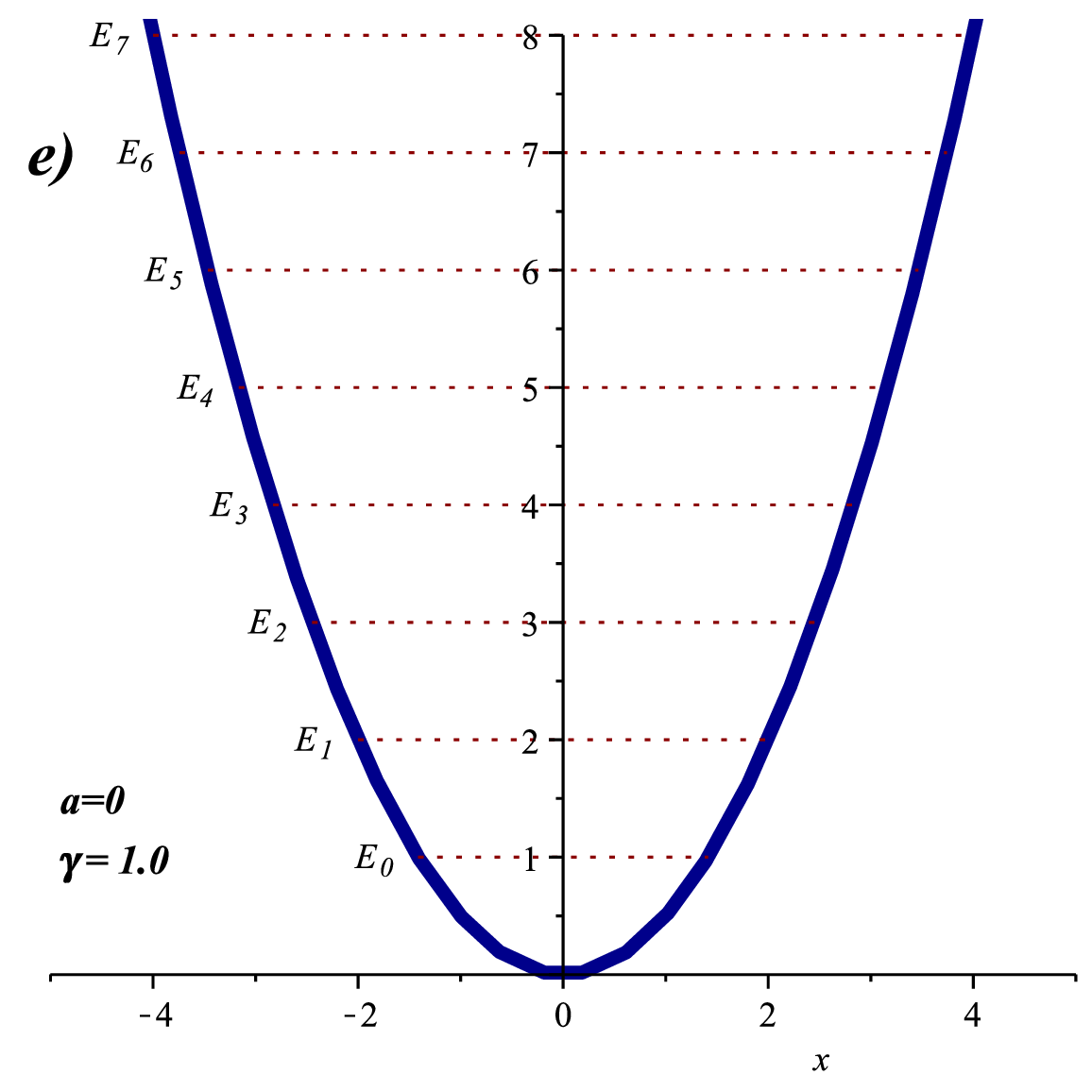}
}
\resizebox{0.32\textwidth}{!}{%
  \includegraphics{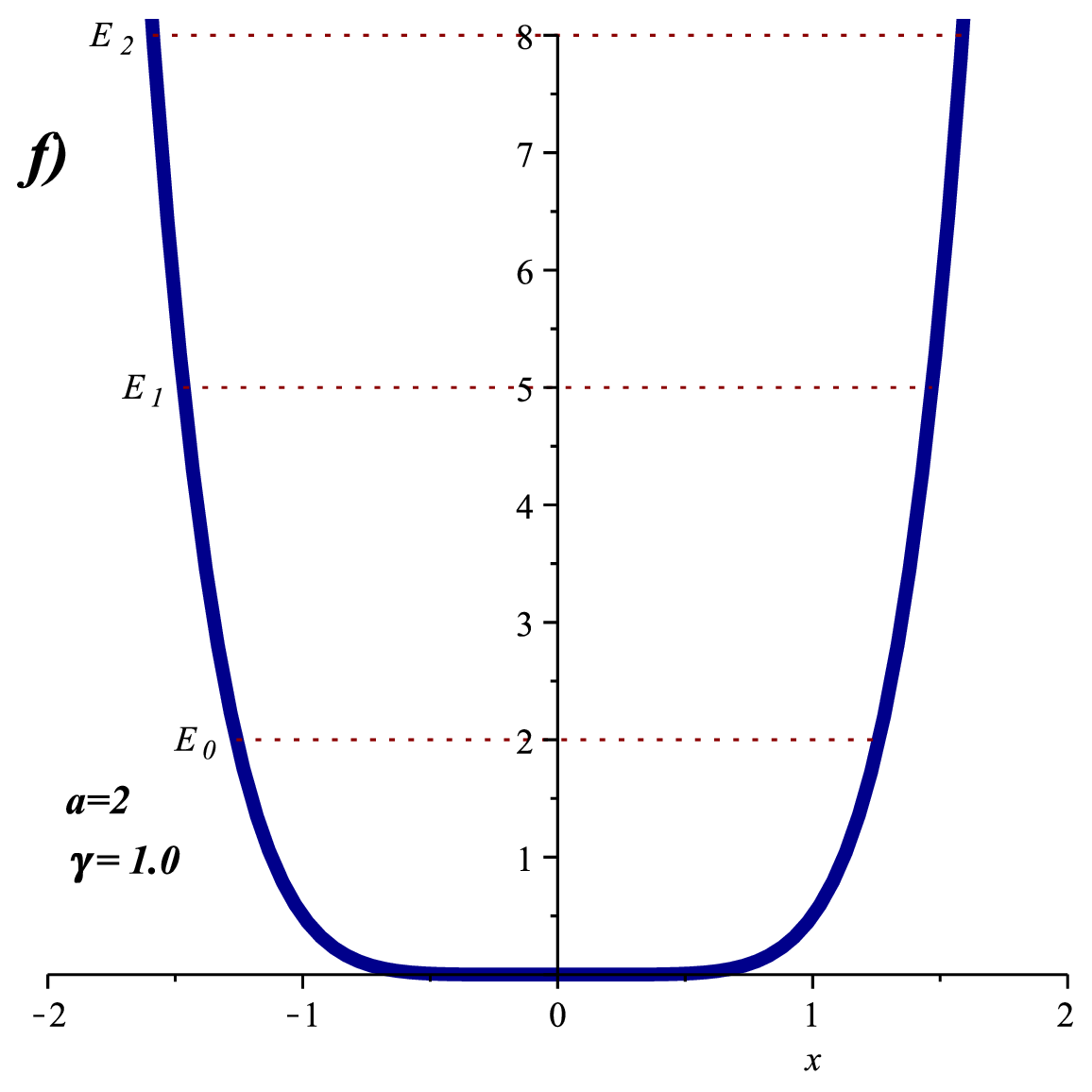}
}\\
\resizebox{0.32\textwidth}{!}{%
  \includegraphics{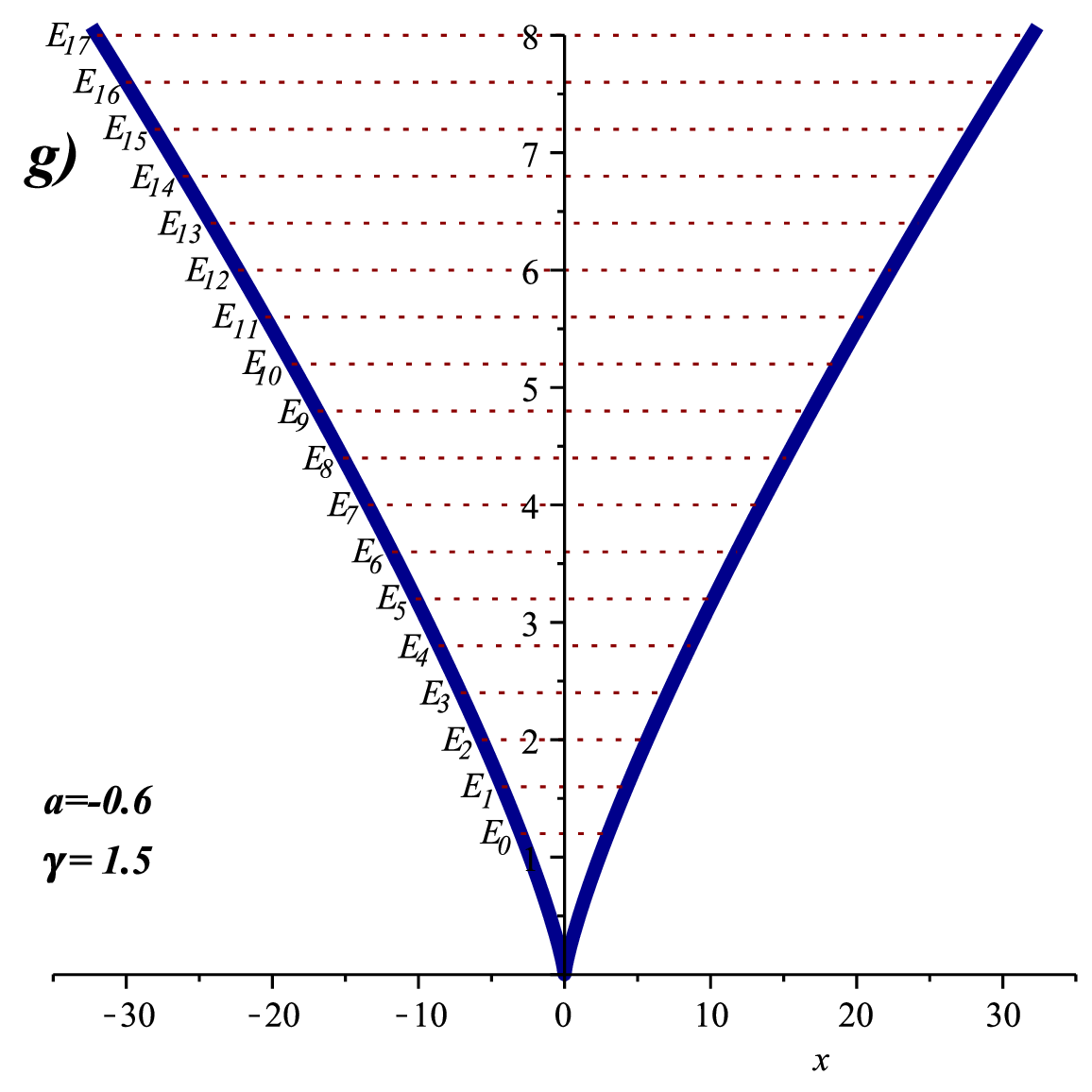}
}
\resizebox{0.32\textwidth}{!}{%
  \includegraphics{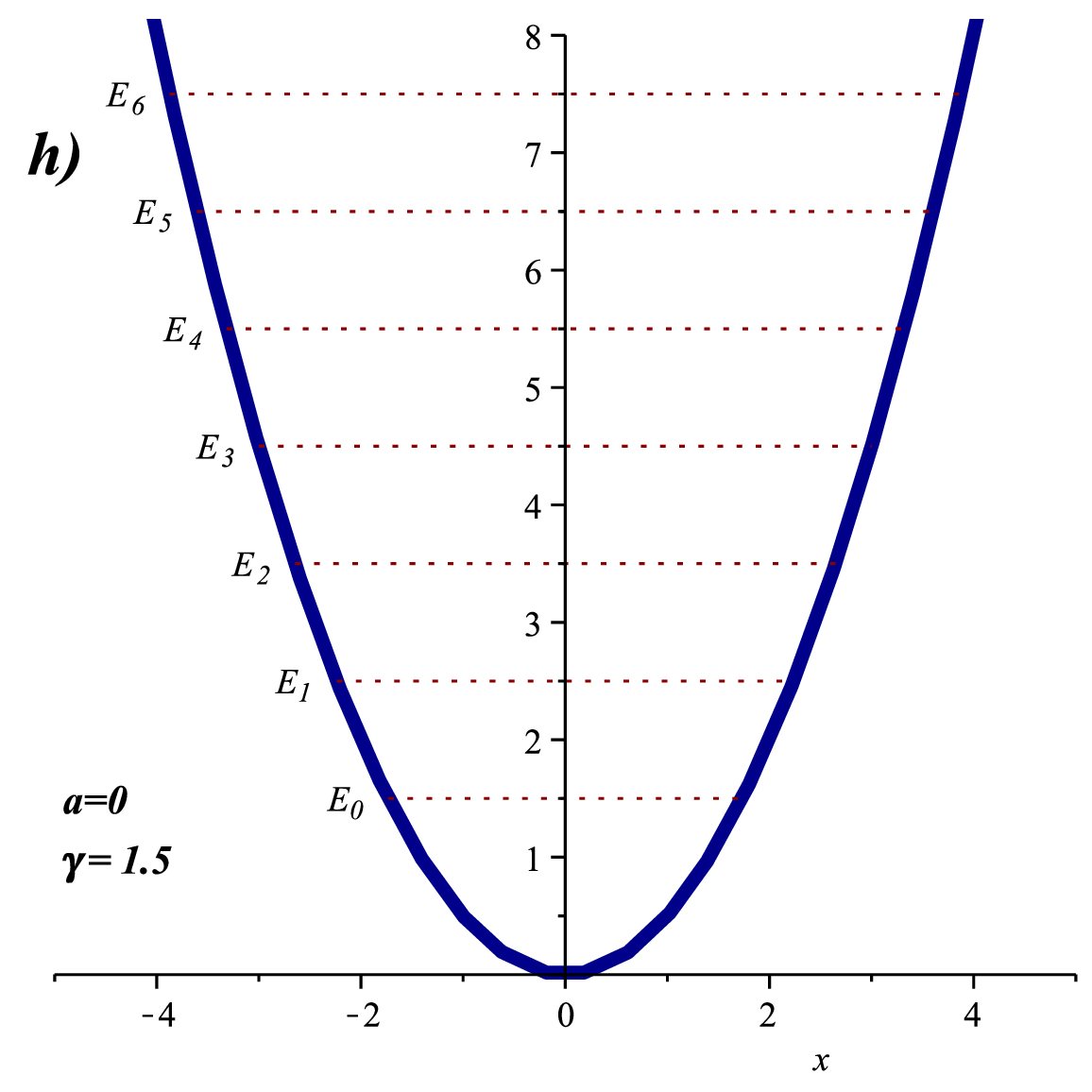}
}
\resizebox{0.32\textwidth}{!}{%
  \includegraphics{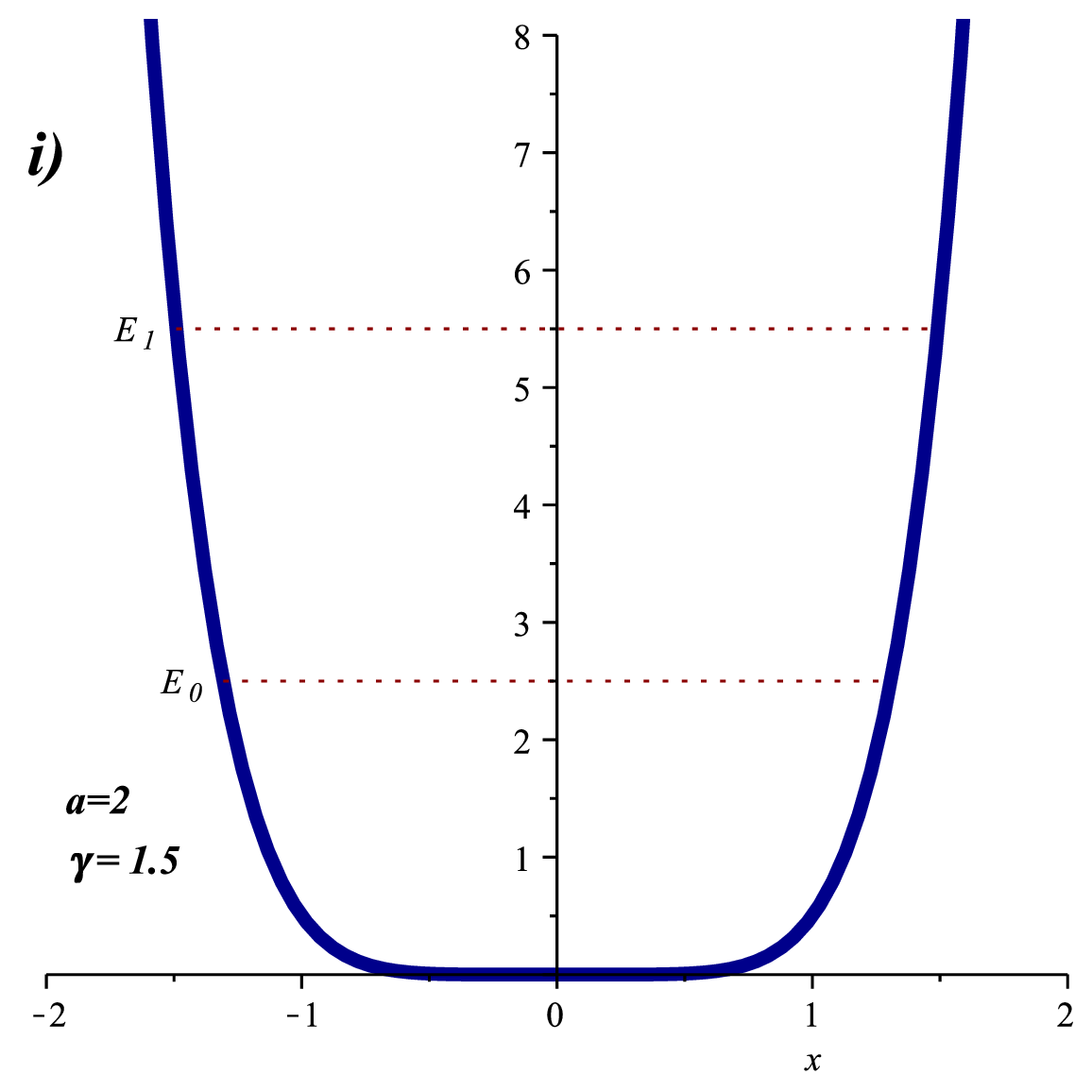}
}
\end{center}
\caption{Quantum harmonic oscillator potential~\eqref{osc-pot} with position-dependent effective mass $M(x)$~\eqref{m-pd} and behavior of the corresponding equidistant energy levels~\eqref{e-kappa}, \eqref{e-2m-alpha2} and \eqref{e-2m1-alpha2} for values of the deformation parameter $a=-0.6;$ $0;$ $2.0$. Upper plots correspond to the parameter $\gamma=0.5$;  middle plots correspond to the parameter $\gamma=1.0$; lower plots correspond to the parameter $\gamma=1.5$ ($m_0=\omega=\hbar=1$).} 
\label{fig.1}
\end{figure}

Before discussing our plots presented in figure~\ref{fig.1}, we want to highlight briefly the results achieved in the preceding part of the current paper. 
In the previous sections, we introduced two propositions on the deformation of the quantum harmonic oscillator Heisenberg-Weyl algebra and its parabose oscillator algebra (or Lie superalgebra $\mathfrak{osp}\left( {1|2} \right)$). 
Both propositions have been proven by straightforward calculations of the relevant commutation relations. 
The parameter $a$ leading to the deformation of the algebras appears due to the generalization of the constant mass to a mass varying with position. 
Next, we have shown that both the algebra and superalgebra deformations can still be realized in terms of the exact solutions of the corresponding quantum oscillator problems.
The realization of such a parameter deformation of the Heisenberg-Weyl algebra is given in Section 3: it exhibits an equidistant energy spectrum of general 
behavior~\eqref{e-kappa}, and the wavefunctions of the stationary states \eqref{psi-fin} corresponding to this energy spectrum are expressed through the Hermite polynomials. 
We demonstrated that a realization of the corresponding deformation of the Lie superalgebra $\mathfrak{osp}\left( {1|2} \right)$ also exists. 
Such a realization is presented in Section 5. 
Its energy spectrum is still equidistant, however, both even and odd state wavefunctions of the parabose oscillator model with the position-dependent mass now are expressed through the Laguerre polynomials. 
In fig.\ref{fig.1}, we depicted the quantum harmonic oscillator potential~\eqref{osc-pot} with position-dependent effective mass $M(x)$~\eqref{m-pd} and the behavior of the corresponding equidistant energy levels~\eqref{e-kappa}, \eqref{e-2m-alpha2} and \eqref{e-2m1-alpha2} for values of the deformation parameter $a=-0.6;$ $0;$ $2.0$. 
The upper plots correspond to the parameter $\gamma=0.5$, the middle plots correspond to the parameter $\gamma=1.0$ and the lower plots correspond to the parameter $\gamma=1.5$. 
For simplicity, the measurement system  $m_0=\omega=\hbar=1$ has been selected for these plots.

First of all, one observes the preservation of the equidistance property for the energy levels despite the fact that the system is now deformed. 
Therefore, different values of the parabose parameter $\gamma$ only change the ground state of the oscillator model, despite the situation that the mass is not constant but changing with position. 
It can be seen from the comparison of the three plots located in each column. 
It is exactly this property that was thoroughly discussed and led to the kernel of the parabose generalization idea of Wigner in~\cite{wigner1950}.

Let us remind that the exact solution for the non-relativistic quantum harmonic oscillator in the canonical approach in terms of the wavefunctions of the stationary states has the following analytical expression in terms of the Hermite polynomials~\cite{landau1991}:

\be
\label{wf-on}
\psi _n \left( x \right) = \frac{1}{{\sqrt {2^n n!} }}\left( {\frac{{m_0\omega }}{{\pi \hbar }}} \right)^{{\frac 14}} e^{ - \frac{{m_0\omega x^2 }}{{2\hbar }}} H_n \left( {\sqrt {\frac{{m_0\omega }}{\hbar }} x} \right).
\ee 

Then, the wavefunctions of the canonical oscillator model with the position-dependent mass \eqref{psi-fin} easily recover this expression for $a=0$.
Furthermore, putting $a=0$ and $\gamma=1/2$ in the even and odd state wavefunctions of the parabose oscillator model with the position-dependent mass also returns \eqref{wf-on}. 
Thus, the energy spectrum expressions \eqref{e-kappa}, \eqref{e-2m-alpha2} and \eqref{e-2m1-alpha2} also easily recover their canonical analogue \eqref{e-c} under the same values of $a$ and $\gamma$. 
This can be observed in the plot (b) of fig.\ref{fig.1}. 
Additionally, plots (e) and (h) exhibit the energy spectrum of the parabose oscillator with a constant mass \eqref{e-nc} under the conditions $a=0$ and $\gamma > 1/2$. 

The first and third columns of plots in fig.\ref{fig.1} also exhibit another important property of the constructed oscillator models. 
It is related to the `hidden' behavior of the deformation parameter $a$ that is included in the model through the mass changing with position. The decrease of the negative value of the deformation parameter $a$ changes the harmonic oscillator potential profile to the triangular-shaped one, if $a \leq -1/2$. This can be observed in the plots (a), (d) and (g) of fig.\ref{fig.1}.

The increase of the value of this deformation parameter shortens the width of the harmonic oscillator potential, and the limit $a \to \infty$ transfers both edges of the harmonic oscillator potential to the potential having a profile very similar to the potential well problem around the oscillator equilibrium point $x=0$. 
This can be considered as a specific confinement effect that appears as a result of the introduction of the position-dependent mass concept to the quantum oscillator system and drastically differs from the oscillator models with a position-dependent mass also exhibiting confinement effects~\cite{jafarov2021,jafarov2022,jafarov2023,jafarov2020,jafarov2020a,jafarov2020c,jafarov2021a,quesne2022}. 
This difference, exact solubility, simple energy spectrum, and uniqueness of the discussed quantum systems exhibiting a specific confinement effect gives great hope for attractive future applications in the field of solid-state physics, nanotechnologies as well as related areas.

\section*{Acknowledgments}

E.I.J. thanks Dr. Loredana Maria Massaro from Eindhoven University of Technology (The Netherlands) for her comments and helpful suggestions, which gave birth to some ideas during the implementation of this work.

\section*{Data Availability Statement}

No data associated in the manuscript.

\end{document}